\documentstyle[12pt]{article}

\renewenvironment{thebibliography}[1]
    {\begin{list}{\arabic{enumi}.}
    {\usecounter{enumi}\setlength{\parsep}{0pt}
\setlength{\leftmargin 1.25cm}{\rightmargin 0pt}
     \setlength{\itemsep}{0pt} \settowidth
    {\labelwidth}{#1.}\sloppy}}{\end{list}}

\begin{document}

\def\be{\begin{equation}}
\def\ee{\end{equation}}
\def\bea{\begin{eqnarray}}
\def\eea{\end{eqnarray}}
\def\CPbar{\hbox{{\rm CP}\hskip-1.80em{/}}}
\def\D0{D\O~}
\def\pbarp{ \bar{{\rm p}} {\rm p} }
\def\pp{ {\rm p} {\rm p} }
\def\ifb{ {\rm fb}^{-1} }
\def\del{\partial }
\def\ra{\rightarrow}

\setcounter{footnote}{1}
\renewcommand{\thefootnote}{\fnsymbol{footnote}}

\begin{titlepage}

hep-ph/9604309 \hfill {\small DESY-95-252; TUIMP-TH-96/75; MSUHEP-60405} 
\vspace{-0.3cm}

\begin{flushright}
{ March, 1996}
\end{flushright}
\vspace*{1.2cm} 
\centerline{\large\bf  
Global Power Counting Analysis On Probing Electroweak}
\baselineskip=18pt
\centerline{\large\bf  
Symmetry Breaking Mechanism At High Energy Colliders   }

\vspace*{1.2cm}
\baselineskip=17pt
\centerline{\normalsize  
{\bf Hong-Jian He}~$^{a}$,
       ~~~ {\bf Yu-Ping Kuang}~$^{b}$,
       ~~~ {\bf C.--P. Yuan}~$^{c}$ } 
 
\vspace*{0.4cm}
\centerline{\normalsize\it
$^{a}$ Theory Division, DESY, D-22603 Hamburg, Germany    }
\vspace{0.2cm}
\centerline{\normalsize\it
$^{b}$ CCAST ( World Laboratory ), 
            P.O.Box 8730, Beijing 100080, China} 
\centerline{\normalsize\it
Institute of Modern Physics, Tsinghua University, 
Beijing 100084, China~\footnote{Mailing address.}}
\vspace{0.2cm}
\centerline{\normalsize\it
$^{c}$
Department of Physics and Astronomy, Michigan State University }
\centerline{\normalsize\it
East Lansing, Michigan 48824 , USA }

\vspace{0.8cm}

\vspace{0.5cm}
\begin{abstract}
\noindent
We develop a precise power counting rule (a generalization of 
Weinberg's counting method for the nonlinear sigma model) 
for the electroweak theories formulated by chiral Lagrangians. 
Then we estimate the contributions of {\it all} next-to-leading order 
(NLO) bosonic operators to the amplitudes of 
the relevant scattering processes which can be measured at 
high energy colliders, such as the LHC and future Linear Colliders.
Based upon these results, we globally classify the sensitivities of 
testing all NLO bosonic operators for probing the electroweak symmetry 
breaking mechanism at high energy colliders.

\end{abstract}

\vspace{1.0cm}
\noindent
PACS number(s): 11.30.Qc, 11.15.Ex, 12.15.Ji, 14.70.-e

\vspace{0.5cm}
\begin{center}
( Published in Physics Letters {\bf B382} (1996) 149-156 )
\end{center}

\end{titlepage}


\newpage
\renewcommand{\thefootnote}{\alph{footnote}}
\setcounter{footnote}{0}
\baselineskip=18pt 

\vspace{0.3cm}
\noindent
{\bf 1. Effective Lagrangian for Strongly Interacting EWSB Sector }
\indent
\vspace{0.15cm}

The current low energy data are sensitive to the 
$SU(2)_L\times U(1)_Y$ gauge interactions of the Standard Model (SM), 
but still allow a wide mass-range (~$65.2~$GeV$\sim O(1)~$TeV~)
for the SM Higgs boson~\cite{LEP2} so that
the electroweak symmetry breaking (EWSB) mechanism remains 
an open question.
The light resonance(s) originating from the EWSB sector with mass(es) 
well below the TeV scale can exist possibly in the SM and necessarily in 
its supersymmetric extensions.
In such cases,  these particles should be detected~\cite{wwlhc,wwnlc}
at the high energy colliders such as the CERN Large Hadron Collider (LHC) 
and the future electron (and photon) Linear Colliders (LC)~\cite{lcws93},
even though the current direct experimental searches so far are all negative.
If the EWSB is, however, driven by strong interactions with no new 
resonance well below the TeV scale,  then it will be a greater
challenge to future colliders to decisively probe the EWSB mechanism. 
This latter case is what we shall study in this work.

It is known that below the scale of any new heavy resonance 
the electroweak chiral Lagrangian (EWCL) provides the most economical 
method to describe the new physics effects, and is 
one of the most important applications of the general idea about 
effective field theories~\cite{eff}.
Following Ref.~\cite{app,etlhc}, the EWCL can be written as
\be
{\cal L}_{eff}
= \displaystyle\sum_n 
\ell_n\displaystyle\frac{f_\pi~^{r_n}}{\Lambda^{a_n}}
{\cal O}_n(W_{\mu\nu},B_{\mu\nu},D_\mu U,U,f,\bar{f})
= {\cal L}_G + {\cal L}_{S} + {\cal L}_F \, ,
\label{eq:effL}
\ee
where 
$D_{\mu}U  =  
\partial_{\mu}U + ig{\bf W}_{\mu}U -ig^{\prime}U{\bf B}_{\mu}~$,
${\bf W}_{\mu}\equiv W^a_{\mu}\displaystyle\frac{\tau^a}{2}$,
${\bf B}_{\mu}\equiv B_{\mu}\displaystyle\frac{\tau^3}{2}$, 
$~U  =  \exp [i\tau^a\pi^a/f_\pi ]~$, $\pi^a$ is the Goldstone boson
(GB) field and $f$($\bar{f}$) is the fermion field.
In (\ref{eq:effL}), we have factorized out the dependence on 
$~f_\pi~$ and $~\Lambda~$ so that the dimensionless 
coefficient $\ell_n$  of the operators ${\cal O}_n$ are
of $O(1)$~~\cite{georgi}.
$f_\pi =246$\,GeV is the vacuum expectation value which 
characterizes the EWSB breaking scale.
The effective cut-off scale $\Lambda$ is the 
highest energy scale below which 
(\ref{eq:effL}) is valid. In the case with no new light 
resonance in the EWSB sector,  
$~\Lambda \approx 4\pi f_\pi~$~\cite{georgi}.
$~{\cal L}_F~$ is the fermionic part of ${\cal L}_{eff}$.\footnote{
Here we concentrate on probing new physics from all possible bosonic
operators and do not include the next-to-leading order fermionic operators
in $~{\cal L}_F~$.}    The bosonic part of the EWCL is given by 
$~{\cal L}_G+{\cal L}_{S}~$ where
${\cal L}_G = -\frac{1}{2}{\rm Tr}({\bf W}_{\mu\nu}{\bf W}^{\mu\nu})
              -\frac{1}{4}B_{\mu\nu}B^{\mu\nu} $ and
$~{\cal L}_S~$ contains operators describing the gauge-boson-GB interactions 
and the GB self-interactions:
\be
{\cal L}_{S}  
 ~=~  {\cal L}^{(2)}+{\cal L}^{(2)\prime}+
             \sum_{n=1}^{14} {\cal L}_n ~.
\label{eq:effL3}
\ee
${\cal L}^{(2)}$ is the universal leading order bosonic operator,
and equals to 
$\frac{f_\pi^2}{4}{\rm Tr}[(D_{\mu}U)^\dagger(D^{\mu}U)]$.
All the other 15 next-to-leading-order (NLO) bosonic operators were 
explicitly given in Refs.~\cite{app,etlhc},
among which
twelve (${\cal L}^{(2)\prime}$ and ${\cal L}_{1\sim 11}$) 
are $CP$-conserving and three 
(${\cal L}_{12\sim 14}$) are $CP$-violating.
Furthermore, the operators $~{\cal L}_{6,7,10}~$ 
violate custodial $SU(2)_C$ symmetry 
(even after $g^{\prime}$ is turned off) in contrast to 
the operators ${\cal L}_{4,5}$ in which the  pure GB interactions
are $SU(2)_C$-invariant.

The coefficients ($\ell_n$'s) of the 15 NLO operators depend on 
the details of the underlying dynamics and reflect the possible
new physics.
Among the 15 NLO coefficients,
$\ell_1$, $\ell_0$ and $\ell_8$ correspond to 
${\rm S}$, ${\rm T}$ and ${\rm U}$
parameters~\cite{app}. ($~{\rm S}=-\ell_1 / \pi$,
$ {\rm T} = \ell_0 /(2\pi e^2)$ and ${\rm U}=-\ell_8 / \pi$.)
They have been measured 
from the current low energy LEP/SLC data 
and will be further improved at LEPII and upgraded Tevatron.
To distinguish different models of the EWSB, 
the rest of the $\ell_n$'s  
has to be measured by studying the scattering processes
involving weak gauge bosons. 
What is usually done in the literature is to consider only 
a small subset of these operators at a time.
For instance, 
in Ref.~\cite{wwlhc}, a non-resonant model (called Delay-K model) was 
studied which includes ${\cal L}^{(2)}$ as well as
the NLO operators ~${\cal L}_4$ and ${\cal L}_5$~. 
It was found that 
for the gold-plated mode (i.e. pure leptonic decay mode) 
of $W^\pm W^\pm$, a total number of about 10 signal events is expected
 at the LHC with a 100\,$\ifb$ luminosity after 
imposing relevant kinematic cuts to suppress backgrounds.
In the end of the analysis the ratio of signal to background is about 1.
Another non-resonant model (called LET-CG model), which contains 
only the model-independent operator ${\cal L}^{(2)}$,
was also studied in that paper.
The difference between the predictions of these two models
signals the effects from the NLO operators ${\cal L}_{4,5}$~.
With just a handful events, it requires
higher integrated luminosities to probe these NLO operators and
compare with the model-independent contributions from ${\cal L}^{(2)}$~.
Generally speaking, if one combines measurements from various 
$VV$-modes, it is possible (although not easy) to distinguish 
models of EWSB which effectively include different subsets
of the 15 NLO operators and the 
model-independent operator ${\cal L}^{(2)}$~.  

The important question to ask is:
`` How and to what extent can one measure 
{\it all}~ the NLO coefficients $\ell_n$  at future colliders  
to {\it fully} explore the EWSB sector? ''
To answer this question, as the first step, one should 
{\bf (i)}. find out, for each given NLO operator, whether 
it can be measured via leading and/or sub-leading amplitudes 
of relevant processes at each collider;
{\bf (ii)}. determine whether a given NLO 
operator can be sensitively (or marginally sensitively) probed 
through its contributions to the leading (or sub-leading) 
amplitudes of the relevant scattering process at each given collider;
{\bf (iii)}. determine whether carrying out the above study for various 
high energy colliders can {\it complementarily} cover all 
the 15 NLO operators to probe the strongly interacting EWSB sector.
For abbreviation, the above requirements {\bf (i)}-{\bf (iii)} 
will be referred hereafter as  the `` {\it Minimal Requirements} ''.

To find the relevant scattering processes 
and determine their sensitivities to
a given NLO operator, one has to first know the contributions of this 
operator to the scattering amplitudes under consideration.
Although one can easily realize whether a single scattering 
process is relevant to probing a given NLO operator or not, 
it is non-trivial to classify all relevant processes to every NLO operator
at different high energy colliders, and to further
determine whether each given NLO operator can be
sensitively/marginally sensitively probed by the corresponding 
scattering processes at these colliders.
This would in principle require detailed calculations on the contributions
of these operators to various scattering amplitudes.
In this work, as a first-step global analysis, we shall only
{\it estimate} the contributions of all these NLO operators 
to various scattering processes
by using a power counting method constructed in Sec.~2.
In Sec.~3, we examine the hierarchy structure for the sizes of the 
scattering amplitudes
and define our theoretical criterion for classifying the sensitivities of 
relevant scattering processes to each NLO operator.
These will answer our {\it Minimal Requirements}-{\bf (i)} and -{\bf (ii)}.
Finally, given the above results, we globally and qualitatively 
classify, in Sec.~4, 
the sensitivities of the relevant scattering processes for
probing all the NLO operators at relevant high energy colliders.
This completes our answer to the  {\it Minimal Requirement}-{\bf (iii)}.

\vspace{0.3cm}
\noindent
{\bf 2. A Power Counting Rule for High Energy Scattering Amplitudes}
\indent
\vspace{0.15cm}

To make a systematic analysis on the sensitivity of a scattering 
process for probing the new physics operators 
in~(\ref{eq:effL}), we have to first compute the scattering amplitudes 
contributed by those operators. 
For this purpose, we generalize 
Weinberg's power counting rule for the ungauged nonlinear 
sigma model (NLSM)~\cite{wei} and develop a 
power counting rule for the EWCL to {\it separately} count the 
power dependences on the energy $E$ and all the relevant mass scales.
Weinberg's counting rule was to count the $E$-power
dependence ($D_E$) for a given $~L$-loop level $S$-matrix element 
$~T~$ in the NLSM. To generalize it to the EWCL, we further include 
the gauge bosons, ghost bosons, fermions  and possible 
$v_\mu$-factors associated with external 
weak gauge boson ($V= W^\pm , Z^0$) lines, 
[cf. (\ref{eq:cbv})]. 
After some algebra, we find that for the EWCL and in the energy region 
$~\Lambda > E \gg M_W, m_t~$, 
\be 
D_E = 2L+2+\sum_n {\cal V}_n
\left( d_n+\frac{1}{2}f_n -2\right) -e_v ~~,
\label{eq:de}                       
\ee
where ${\cal V}_n$ is the number of type-$n$ vertices in $T$,
$d_n$($f_n$) is the number of derivatives (fermion-lines) 
at a vertex of type-$n$, and
$e_v$ is the number of possible external $v^\mu$-factors
[c.f. (\ref{eq:cbv})]. 
For external fermions, we consider masses 
$~~m_f\leq m_t \sim O(M_W)\ll E ~$, and the spinor wave functions are 
normalized as 
$~~ \bar{u}(p,s)u(p,s^\prime )=2m_f\delta_{ss^\prime}~~$, etc.

To correctly estimate the magnitude of each given amplitude $~T~$,
besides counting the power of $E$,
it is also crucial to {\it separately} count the
power dependences on the two typical mass scales of the EWCL: 
the vacuum expectation value $f_\pi$ and the effective cut-off 
$\Lambda$ of the effective theory.\footnote{ 
If the powers of $f_\pi$ and $\Lambda$ 
are not separately counted, 
$\Lambda/f_\pi \simeq 4\pi$ will be mistakenly counted as 1. 
This makes the estimated results off by orders of
magnitude. If a power counting rule only counts 
the sum ~$D_E+D_\Lambda$~~\cite{hveltman},
it cannot be used to correctly estimate the order of magnitudes. 
E.g., the amplitudes $\frac{E^2}{f_\pi^2}$ and 
$\frac{E^2}{f_\pi^2}\frac{E^2}{\Lambda^2}$ have the {\it same}
$D_E+D_\Lambda$ but are clearly 
of different orders in magnitude. (For $E=1$~TeV, they
differ by a factor $\sim 10$.)}~
The $\Lambda$-dependence comes from the NLO operator tree-level vertices, 
each of which contributes a factor $~1/\Lambda^{a_n}~$ 
[cf.~(\ref{eq:effL})] so that the total factor is
$~1/\Lambda^{\sum_n a_n}~$.~ The power factor $~\Lambda^{a_n}~$
associated with each operator $~{\cal O}_n~$ can be counted by 
the naive dimensional analysis (NDA)~\cite{georgi}.
In general, ~$T$~ can always be
written as $~f_\pi^{D_T}~$ times some dimensionless function of 
$~E,~\Lambda$ and $f_\pi$, 
where $D_T = 4-e$ and 
$e$ is the number of external bosonic and fermionic lines.  
Bearing in mind the intrinsic $L$-loop factor  
$~(\frac{1}{16\pi^2})^L=(\frac{1}{4\pi})^{2L}~$,  
we can then construct the following precise counting rule for $~T~$
in the energy region $~\Lambda >E \gg M_W, m_t~$:
\bea
T= c_T f_\pi^{D_T}\displaystyle 
\left(\frac{f_\pi}{\Lambda}\right)^{N_{\cal O}}
\left(\frac{E}{f_\pi}\right)^{D_{E0}}
\left(\frac{E}{4 \pi f_\pi}\right)^{D_{EL}}
\left(\frac{M_W}{E}\right)^{e_v} H(\ln E/\mu) ~~,\nonumber \\[0.3cm]
N_{\cal O}=\sum_n a_n~,~~~~ 
\displaystyle D_{E0}=2+\sum_n {\cal V}_n
\left( d_n+\frac{1}{2}f_n-2\right)~, ~~~~ 
D_{EL}=2L~,
\label{eq:counting}
\eea
where the dimensionless coefficient $~c_T~$ contains 
possible powers of gauge couplings ($g,g^\prime$) and Yukawa 
couplings ($y_f$) from the vertices 
of $~T~$, which can be directly counted. $~H$~ is a function 
of $~\ln (E/\mu )~$ coming from loop corrections in the standard
dimensional regularization~\cite{eff}
and is insensitive to $E$.  Neglecting the
insensitive factor $~H(\ln E/\mu)$, we can extract the main features of
scattering amplitudes by simply applying~(\ref{eq:counting}) to
the corresponding Feynman diagrams. 

Note that the counting for $E$-power dependence in 
(\ref{eq:de}) or (\ref{eq:counting})
cannot be directly applied to the amplitudes with external
longitudinal gauge boson ($V_L$) lines. 
Consider the tree-level $V_L V_L \rightarrow V_L V_L$ 
amplitude. Using (\ref{eq:counting}) 
and adding the $E$-factors from the four longitudinal polarization vectors
$~\epsilon_L^{\mu}\sim k^\mu /M_{W,Z}~$, we find that the leading amplitude
is proportional to $E^4/f_\pi^4$ which 
violates the low energy theorem result (i.e. $E^2/f_\pi^2$).
This is because the naive power counting for $V_L$-amplitudes
only gives the leading $E$-power of individual Feynman diagrams, it
does not reflect the fact that gauge invariance causes the 
cancellations of the $E^4$-terms among individual diagrams.
So, how can we count $D_E$ in any amplitude with external $V_L$-lines?
We find that this can be elegantly solved by using the 
Ward-Takahashi (WT) identity 
(cf. \cite{et} for a precise derivation\footnote{
This identity was used to derive the Equivalence Theorem (ET) to all
orders in the perturbative expansion and to explore the
profound physical content of the ET \cite{et}. }):
\be
T[V^{a_1}_L,\cdots ,V^{a_n}_L;\Phi_{\alpha}]
= C\cdot T[-i\pi^{a_1},\cdots ,-i\pi^{a_n};\Phi_{\alpha}]+ B~~,\\
\label{eq:st}
\ee                                               
with\vspace{-0.8cm}
\bea
C \equiv C^{a_1}_{mod}\cdots C^{a_n}_{mod} ~,~~
 v^a  \equiv  v^{\mu}V^a_{\mu}~,~ 
~~v^{\mu}~\equiv \epsilon^{\mu}_L-k^\mu /M_a 
= O(M_a/E) ~, \nonumber \\ 
B \equiv \sum_{l=1}^n \{ ~C^{a_{l+1}}_{mod}\cdots C^{a_n}_{mod}
\,T[v^{a_1},\cdots ,v^{a_l},-i\pi^{a_{l+1}},\cdots ,
 -i\pi^{a_n};\Phi_{\alpha}] +{\rm permutations} ~ \}~,
\label{eq:cbv}
\eea  
where $~\pi^a$   are GB fields and $\Phi_{\alpha}$ denotes other 
possible physical in/out states. The 
constant modification factor
$~C_{mod}^a=1+O({g^2 \over 16 \pi^2})$ in the EWCL
and can be exactly simplified as $1$  
in certain convenient renormalization schemes~\cite{et}. 
Since the right-hand side (RHS)  of (\ref{eq:st}) does not 
have $E$-power cancellations related to external legs,
we can therefore apply our counting rule (\ref{eq:counting}) to 
{\it indirectly count the $D_E$ of
the $V_L$-amplitude via counting the $D_E$ of the RHS of (\ref{eq:st}).}

\vspace{0.3cm}
\noindent 
{\bf 3. Estimating Scattering Amplitudes and Analyzing Their Sensitivities
to Each Given Operator}
\indent
\vspace{0.15cm}

The main advantage of using the above counting rule
(\ref{eq:counting}) is that we can correctly and quickly estimate 
the magnitude of any scattering amplitude
in the energy region $M_{W}, m_t \ll E < \Lambda$.
Using the above counting rule (\ref{eq:counting}), 
we have performed a global analysis for
all $~V^aV^b \ra V^cV^d~$ and $~q\bar{q}\ra V^aV^b~$ processes
by estimating the contributions from both model-independent operator
$~{\cal L}_0$($\equiv {\cal L}_G +{\cal L}_F +{\cal L}^{(2)}~$) 
up to one-loop and the other 15  model-dependent NLO operators
at tree-level~\cite{etlhc}. 
We observe a power counting hierarchy in terms 
of $E$, $f_\pi$ and $\Lambda$ for these amplitudes:
\be
\frac{E^2}{f_\pi^2}\gg \frac{E^2}{f_\pi^2}\frac{E^2}{\Lambda^2}, 
~g\frac{E}{f_\pi} \gg g\frac{E}{f_\pi}\frac{E^2}{\Lambda^2}, ~g^2 
\gg g^2\frac{E^2}{\Lambda^2}, ~g^3\frac{f_\pi}{E} 
\gg g^3\frac{Ef_\pi}{\Lambda^2},~g^4\frac{f_\pi^2}{E^2} 
\gg g^4\frac{f_\pi^2}{\Lambda^2}~,
\label{eq:pch}
\ee
which, in the typical high energy region  
$~~E\in (750\,{\rm GeV},~1.5\,{\rm TeV})$, gives
$$ 
\begin{array}{c}
 (9.3,37)\gg (0.55,8.8),(2.0,4.0)\gg (0.12,0.93),(0.42,0.42) \gg 
     \nonumber \\
(0.025,0.099),(0.089,0.045)\gg 
(5.3,10.5)\times 10^{-3},(19.0,4.7)\times 10^{-3} 
\gg (1.1,1.1)\times 10^{-3} ~~,
\nonumber
\end{array}
$$ 
where $E$ is taken to be the $VV$-pair invariant mass
and $~\Lambda\approx 4\pi f_\pi~$. 
This power counting hierarchy is easy to understand. 
In (\ref{eq:pch}), from left to right, the hierarchy
is built up by increasing either the number of derivatives (i.e. 
power of $E/\Lambda$) or the number of external transverse gauge boson 
$V_T$'s (i.e. the power of gauge couplings).  This power counting
hierarchy provides us a theoretical base to classify all
the relevant scattering amplitudes 
in terms of the three essential parameters $E$, $f_\pi$ and $\Lambda$ 
plus possible gauge/Yukawa coupling constants.
In the high energy region $M_W, m_t\ll E <\Lambda $ and to each order
of chiral perturbation, for a given type of 
processes [which all contain the same number of external $V$-lines
($V=W^\pm , {\rm or}, Z$) with other external lines exactly the same],
the leading amplitude is given by the amplitude
with all external $V$-lines being longitudinal, 
and the sub-leading amplitude is given by
the amplitude with only one external $V_T$-line (and all other
external $V$-lines being longitudinal). This is because 
the EWCL formalism is a momentum-expansion and 
the GBs (and thus $V_L$'s) are derivatively coupled. 

To answer the {\it Minimal Requirement}-{\bf (i)}, 
we classify in Table~2 the most important
leading and sub-leading amplitudes that can probe the
NLO operators via various processes.\footnote{
Other amplitudes below the sub-leading amplitude for each type of
processes are given elsewhere~\cite{etlhc}.}~
To answer the {\it Minimal Requirement}-{\bf (ii)}, we  
shall establish a theoretical criterion for classifying the 
{\it sensitivity} of a given scattering process to each NLO operator.

Let us  consider the scattering process
$W^\pm W^\pm \ra W^\pm W^\pm$ as a typical example to illustrate the idea.
The leading and sub-leading amplitudes for this process are given by 
the one with four external $W_L$-lines
(~$T[4W_L]$~) and the one with three external $W_L$-lines plus one 
$W_T$-line (~$T[3W_L,W_T]$~), respectively.
In Table~1a we 
estimate the tree and one-loop level contributions 
from the model-independent operators in 
$~{\cal L}_0 \equiv {\cal L}_G +{\cal L}_F +{\cal L}^{(2)}~$ 
to the leading amplitude ~$T[4W_L]$~ and to the sub-leading amplitude 
$~T[3W_L,W_T]~$. 
In the same table, we also list the model-independent 
contributions to various $B$-terms [cf. (\ref{eq:cbv})].
($B^{(0)}$ and $B^{(1)}$ 
denote the $B$-term from $V_L$-amplitudes 
with $0$ and $1$ external $V_T$-line, respectively.) 
In Table~1b, we list the tree-level contributions 
from the model-dependent operators to these two amplitudes.
For instance, the model-dependent leading contributions in
$~T[4W_L]~$ come from the operators
$~{\cal L}_{4,5}~$. (The contributions from
$~{\cal L}_{2,3,9}~$ in $~T[4W_L]~$ are suppressed by a factor 
$~E^2/f_\pi^2~$ relative to that from $~{\cal L}_{4,5}~$.)
Therefore, it is easier to measure $~{\cal L}_{4,5}~$ than
$~{\cal L}_{2,3,9}~$ via the 
$W^\pm_L W^\pm_L \ra W^\pm_L W^\pm_L$ process.
From Table~1b, we also learn that the largest contributions in
the sub-leading amplitude $~T[3W_L,W_T]~$ 
come from $~{\cal L}_{3,4,5,9,11,12}~$.
To determine which operators can be sensitively probed via a given 
process, we introduce the following theoretical criterion on the  
sensitivity
of the process to probing a NLO operator.
Consider the contributions of
$~{\cal L}_{4,5}~$ to $~T[4W_L]~$ as an example.
For this case, the WT identity (\ref{eq:st}) gives, 
\be
T[4 W_L]
= C\cdot T[4\pi] + B~~,\\
\label{eq:examp}
\ee                                               
where $~C=1+O({g^2 \over 16 \pi^2})$~, $T[4\pi]=T_0[4\pi]+T_1[4\pi]$
and $B=B^{(0)}_0+B^{(0)}_1$, in which ~$T_1[4\pi]$~
contains both the model-independent [~$E^4/(16 \pi^2 f^4_\pi)~$]
and model-dependent contributions 
[~$\ell_{4,5}E^4/(f^2_\pi \Lambda^2)~$], cf. Table~1a,b. 
Similarly, $~B^{(0)}_1$  
contains both the model-independent [~$g^2E^2/(16 \pi^2f_\pi^2)~$]
and model-dependent 
[~$\ell_{4,5}~g^2 E^2/\Lambda^2~$]  contributions.
Note that the leading $B$-term $B^{(0)}_0$, which is
of $~O(g^2)$, only depends on the SM gauge couplings and
is of the same order as the leading pure 
$W_T$-amplitude $~T[4W_T]~$~\cite{et,etlhc}.
Thus, $B$ {\it is insensitive to the EWSB mechanism}.
To {\it sensitively probe} the EWSB sector by measuring 
$~{\cal L}_{4,5}~$ via $~T[4W_L]~$ amplitude, 
we demand the pure GB-amplitude $~T[4\pi ]~$ contributed from 
$~\ell_{4,5}~$ (as a direct reflection of the EWSB dynamics) 
to dominate over the corresponding model-independent
leading $B$-term (~$B_0^{(0)}~$), i.e. reguiring 
~$\ell_{4,5}E^4/(f^2_\pi \Lambda^2) \gg g^2$.
This gives,
for $~\ell_{4,5}=O(1)~$,   ~$ \frac{1}{4}\frac{E^2}{\Lambda^2}
  \gg \frac{M^2_W}{E^2}~$,
or $~1\gg (0.7 \,{\rm TeV}/E)^4 ~$.\footnote{ 
This condition was first correctly derived in the
1st paper of Ref.~\cite{et} for the EWCL
and is different from that in Ref.~\cite{etdo} where
 the $B$-term was incorrectly estimated as 
$O(M_W/E)$ instead of $O(g^2)$.  
Also, $f_\pi$ and $\Lambda$ were not separately 
counted for $T_0$ and  $T_1$ in Ref.~\cite{etdo} so that
the factor $\frac{\Lambda^2}{f_\pi^2}$
($\approx 16\pi^2 \geq O(10^2)$) was mistaken as $1$.  After
private communications, the authors of Ref.~\cite{etdo} informed us
that they agreed with our condition (see footnote-20 in the 1st paper of
Ref.~\cite{et}).}~
Thus, {\it sensitively probing}  $~{\cal L}_{4,5}~$ 
via the $~4W^\pm_L$-process requires $E\geq 1$\,TeV, 
which agrees with the conclusion from
a detailed Monte Carlo study in Ref.~\cite{wwlhc}.

It is straightforward
to generalize the above discussion to any scattering 
process up to the NLO. In this paper, we generally classify the sensitivities
of the processes as follows.
For a scattering process involving the NLO coefficient $\ell_n$, if 
$T_1 \gg B~$, then this process is classified to be {\it sensitive} to the 
operator ${\cal L}_n$~.
If not, this process is classified to be 
either {\it marginally sensitive} (for $~T_1 > B~$ but $~T_1 \not\gg B~$)
or {\it insensitive} (for $~T_1 \leq B~$) to the operator ${\cal L}_n$. 
In Tables~1 and 2, {\it both the GB-amplitude and the $B$-term 
are explicitly estimated by our counting rule (\ref{eq:counting}).}
 If $~T_1\leq B~$, this means that the sensitivity is poor so that the
probe of $T_1$ is experimentally harder and requires a higher experimental
precision of at least $O(B)$ to test $T_1$. 
The issue of whether to numerically include $B$
in an explicit calculation of the $V_L$-amplitude is {\it irrelevant}
to the above conclusion.

\vspace{0.3cm}
\noindent
{\bf 4. Classification of Sensitivities to Probing EWSB Sector at 
        Future High Energy Colliders }
\indent
\vspace{0.15cm}

This section is devoted to discuss our 
{\it Minimal Requirement}-{\bf (iii)}.
It is understood that the actual sensitivity
of a collider to probe the NLO operators depends
not only on the luminosities of the active partons (including 
weak-gauge bosons)
inside hadrons or electrons (as discussed in Ref.~\cite{etlhc}),
but also on the detection efficiency of the signal events after
applying background-suppressing kinematic cuts
to observe the specific decay mode of the final state
weak-bosons (as discussed in Refs.~\cite{wwlhc,wwnlc}). 
However, all of these will only add fine structures 
to the sub-leading contributions listed in Table~2 but not 
affect our conclusions about the leading contributions
as long as there are enough signal events produced.
This fact was illustrated in Ref.~\cite{etlhc}
for probing the NLO operators via 
$W^\pm W^\pm \ra W^\pm W^\pm$ at the LHC. 
We have further applied the same method to 
other scattering processes (including possible incoming
photon/fermion fields) 
for various high energy colliders with the luminosities of
the active partons included, 
the details of the study will be given elsewhere.
In this paper, we shall not perform a detailed numerical study 
like Refs.~\cite{wwlhc,wwnlc}, 
but only give a first-step qualitative global power counting 
analysis which serves as a useful guideline for further elaborating
numerical calculations.

After examining all the relevant $ 2 \ra 2$ and $2 \ra 3$ hard scattering
processes, we summarize in Table~2 our global classification for 
the sensitivities of various 
future high energy colliders to probing the 15 
model-dependent NLO bosonic operators.
Here, the energy-$E$ represents the typical energy scale
of the hard scattering processes under consideration.
The leading $B$-term for each high energy
process is also listed and compared with the corresponding
$V_L$-amplitude. If the polarizations of the 
initial/final state gauge bosons are not distinguished 
but simply summed up, the largest $B$ 
in each process (including all possible polarization states)
should be considered for comparison. 
[If the leading $B_0$ (with just one $v_\mu$-factor, cf. eq.~(6))
 happens to be zero, then the largest next-to-leading
 term, either the part of $B_0$ term that contains
$2$ (or $3$) $v_\mu$-factors or the $B_1$ term, should be considered.
Examples are the $~ZZ\rightarrow ZZ~$ and
$~f\bar{f}\rightarrow ZZZ~$ processes.]
By comparing $T_1$ with $B$ in Table~2 and applying
our criterion for classifying the sensitivities, we find that
for the typical energy scale ($E$) of the relevant processes at each
collider, the leading contributions (~marked by $\surd~$) 
can be sensitively probed, while the sub-leading contributions
(~marked by $\triangle~$) can only be marginally sensitively 
probed.\footnote{The exceptions are 
$~f\bar{f}^{(\prime )}\rightarrow W^+W^-/(LT),W^\pm Z/(LT)~$ 
for which $~T_1 \leq B_0$~. Thus the probe of them is insensitive.
($L/T$ denotes the longitudinal/transverse polarizations of
$~W^\pm ,~Z^0~$ bosons.) }
(To save space, 
Table~2 does not list those processes to which the NLO operators 
{\it only} contribute sub-leading amplitudes. These processes
are $~WW\rightarrow W\gamma ,Z\gamma +{\rm perm.}~$ 
and $~f\bar{f}^{(\prime )}\rightarrow W\gamma ,WW\gamma , WZ\gamma 
~$, which all have one external transverse $\gamma$-line and 
are at most marginally sensitive.)

From Table~2, some of our conclusions can be drawn as follows. \\
~~~{\bf (1).}
At LC(0.5), which is a LC with $\sqrt{S}=0.5$\,TeV, $\ell_{2,3,9}$
can be sensitively probed via $e^-e^+ \ra W^-_L W^+_L$. \\ 
~~~{\bf (2).}
For pure $V_L V_L \ra V_L V_L$ scattering amplitudes, 
the model-dependent operators ${\cal L}_{4,5}$
and ${\cal L}_{6,7}$ can be probed 
most sensitively. 
 ${\ell}_{10}$ can only be sensitively probed 
via the scattering process $Z_LZ_L \ra Z_LZ_L$ which 
is easier to detect at the LC(1.5) [a $e^-e^+$ or $e^-e^-$ collider 
with $\sqrt{S}=1.5$\,TeV] than at the LHC(14) [a pp collider with
$\sqrt{S}=14$\,TeV]. \\ 
~~~{\bf (3).}
The contributions from ${\cal L}^{(2)\prime}$~ and 
${\cal L}_{2,3,9}$ to the pure $4V_L$-scattering processes
 lose the $E$-power dependence by a 
factor of $2$ (see, e.g., Table~1b). Hence, the pure $4V_L$-channel is 
less sensitive to these operators. 
[Note that ${\cal L}_{2,3,9}$  can be sensitively 
probed via $f {\bar f} \ra W_L^-W_L^+$ process at LC(0.5) and LHC(14).]
The pure $4V_L$-channel cannot probe ${\cal L}_{1,8,11\sim 14}$ which
can only be probed via processes with $V_T$('s). 
Among ${\cal L}_{1,8,11\sim 14}$,
the contributions from $~{\cal L}_{11,12}~$ to processes 
with $V_T$('s) are most important, although their contributions 
are relatively suppressed by a factor $gf_\pi /E$  as compared to
the leading contributions from 
${\cal L}_{4,5}$ to pure $4V_L$-scatterings.
${\cal L}_{1,8,13,14}$ are generally suppressed by higher powers of
$gf_\pi /E$ and are thus the least sensitive.
The above conclusions hold for both LHC(14) 
and LC(1.5). \\
~~~{\bf (4).} 
At LHC(14), ${\ell}_{11,12}$ 
can be sensitively probed via $q \bar q' \ra W^\pm Z$
whose final state is not electrically neutral. Thus, 
this final state is not accessible at LC. 
Hence, LC(0.5) will not be sensitive to these operators.
To sensitively probe ${\ell}_{11,12}$ at LC(1.5), one has to measure
$e^-e^+ \ra W^-_L W^+_L Z_L$. 
\\ 
~~~{\bf (5).}
To sensitively probe ${\ell}_{13,14}$,
a high energy $~e^-\gamma~$ linear collider 
is needed for studying the processes 
$~e^-\gamma \ra \nu_e W^-_LZ_L,~e^-W^-_LW^+_L~$, 
in which the backgrounds \cite{eAback}~ are much
less severe than processes like $\gamma \gamma \ra W^+_L W^-_L$ at
a $\gamma\gamma$ collider~\cite{DPF,lcws93}.\footnote{The amplitude of
$\gamma \gamma \ra W^+_L W^-_L$ has the order of 
$~e^2\frac{E^2}{\Lambda^2}~$, to which the ${\cal L}_{13,14}$
(and also ${\cal L}_{1,2,3,8,9}$) can contribute. Thus, this process
would be useful for probing ${\ell}_{13,14}$ 
at a $\gamma\gamma$ collider
if the backgrounds somehow could be efficiently suppressed.}

From the above global analysis,  we speculate\footnote{
To further reach a detailed quantitative conclusion, 
an elaborate and precise numerical study on all signal/background rates 
is necessary.}~ that 
before having a large number of signal events at the LHC (i.e. with 
large integrated luminosity), the LHC alone will not be 
able to sensitively measure all these operators,
the linear collider is needed to {\it complementarily}
cover the rest of the NLO operators.
In fact, the different phases of 500 GeV and 1.5 TeV energies at the LC
are necessary because they will be sensitive to different 
NLO operators in the EWCL.  An electron-photon (or a photon-photon) 
collider is also very useful for measuring
all the NLO operators which distinguish different models
of the EWSB in the strongly interacting scenario. 

\vspace{0.3cm}
\noindent
{\bf Acknowledgments}
~~~~We thank Mike Chanowitz, Tao Han, Francisco Larios, and Peter Zerwas 
for reading the manuscript and for useful suggestions. 
H.J.H. is supported by the AvH of Germany and the U.S. DOE; 
Y.P.K. by the NSF of China and the Tsinghua FRF; and
C.P.Y. by the U.S. NSF.


\newpage
\addtolength{\topmargin}{0.5in}
\addtolength{\textwidth} {-0.8in}
\addtolength{\oddsidemargin} {0.4in}
\addtolength{\evensidemargin}{0.4in} 
\setlength{\leftmargin 1.00cm}{\rightmargin 0pt}
\renewcommand{\baselinestretch}{1.3}
\addtolength{\textwidth}{3.8cm}
\addtolength{\oddsidemargin}{-2.4cm}
\evensidemargin=\oddsidemargin
\addtolength{\textheight}{3.6cm}
\addtolength{\topmargin}{-2.5cm}
\parskip=5pt plus 1pt minus 1pt
\renewcommand{\baselinestretch}{1.8}

\begin{table}[t]  
\begin{center}

{\bf Table 1.}  Estimates of leading/sub-leading amplitudes and the
corresponding $B$-terms for $~W^{\pm}W^{\pm}\rightarrow W^{\pm}W^{\pm}~$
scatterings. 
\vspace{0.3cm}

{\bf Table 1a.} 
~Model-independent contributions from 
$~{\cal L}_G+{\cal L}_F +{\cal L}^{(2)}~$~ (~$\Lambda_0\equiv 4\pi f_\pi~$).
\vspace{0.5cm}

\small

\begin{tabular}{||c||c|c||c|c||} 
\hline\hline
${\cal L}_G +{\cal L}_F + {\cal L}^{(2)}$  
&  $~~~T_{\ell}[4\pi]~~~     $  
&  $~T_{\ell}[3\pi,W_T]~ $  
&  $~~~B_{\ell}^{(0)}~~~     $  
&  $~~~B_{\ell}^{(1)}~~~     $ \\   
\hline\hline
Tree-Level~(${\ell}=0$)
&  $ \frac{E^2}{f_{\pi}^2} $
&  $ g\frac{E}{f_\pi} $
&  $ g^2$
&  $ g^2\frac{M_W}{E} $  \\
\hline
One-Loop~(${\ell}=1$)  
& $\frac{E^2}{f_{\pi}^2}\frac{E^2}{\Lambda_0^2}$ 
& $g\frac{E}{f_{\pi}}\frac{E^2}{\Lambda_0^2}$ 
& $g^2\frac{E^2}{\Lambda_0^2}$ 
& $g^3\frac{Ef_\pi}{\Lambda_0^2}$ \\
\hline\hline 
\end{tabular}
\end{center}
\end{table}

\vspace{0.8cm}


\begin{table}[t]  
\begin{center}

{\bf Table 1b.} Model-dependent contributions from each NLO operator. \\ 
(Note: Up to the order $1/\Lambda^2$,
${\cal L}_{6,7,10}$ do not contribute to this process.)

\vspace{0.8cm}

\renewcommand{\baselinestretch}{0.16}
\footnotesize
\begin{tabular}{||c||c|c|c|c|c|c|c|c||} 
\hline\hline
& & & & & & & &  \\
Operators 
& $ {\cal L}^{(2)\prime} $ 
& $ {\cal L}_{1,13} $ 
& $ {\cal L}_2 $
& $ {\cal L}_3 $
& $ {\cal L}_{4,5} $
& $ {\cal L}_{8,14} $ 
& $ {\cal L}_{9} $ 
& $ {\cal L}_{11,12} $ \\
& & & & & & & &  \\
\hline\hline
& & & & & & & &  \\
$ T_1[4\pi] $ 
& $\ell_0 ~\frac{E^2}{\Lambda^2}$
& /
&  $\ell_2 ~e^{2}\frac{E^2}{\Lambda^2}$  
& $\ell_3 ~g^2\frac{E^2}{\Lambda^2}$  
& $\ell_{4,5} ~\frac{E^2}{f_{\pi}^2}\frac{E^2}{\Lambda^2}$  
& /
& $\ell_9 ~g^2\frac{E^2}{\Lambda^2}$
& /  \\
& & & & & & & &  \\
\hline
& & & & & & & &  \\
  $ T_1[3\pi , W_T] $ 
& $\ell_0 ~g\frac{f_\pi E}{\Lambda^2}$
& $\ell_{1,13} ~e^2g\frac{f_{\pi}E}{\Lambda^2}$
& $\ell_2 ~e^2g\frac{f_{\pi}E}{\Lambda^2}$ 
& $\ell_3 ~g\frac{E}{f_{\pi}}\frac{E^2}{\Lambda^2}$ 
& $\ell_{4,5} ~g\frac{E}{f_{\pi}}\frac{E^2}{\Lambda^2}$ 
& $\ell_{8,14} ~g^3\frac{f_\pi E}{\Lambda^2}$ 
& $\ell_9 ~g\frac{E}{f_{\pi}}\frac{E^2}{\Lambda^2}$ 
& $\ell_{11,12}~ g\frac{E}{f_{\pi}}\frac{E^2}{\Lambda^2}$ \\
& & & & & & & &  \\
\hline\hline
& & & & & & & &  \\
 $B_1^{(0)}$
& $\ell_0~       g^2\frac{f^2_{\pi}}{\Lambda^2}$
& $\ell_{1,13}~  e^2g^2\frac{f^2_{\pi}}{\Lambda^2}$
& $\ell_2~       e^2g^2\frac{f_\pi^2}{\Lambda^2}$
& $\ell_3~       g^2\frac{E^2}{\Lambda^2}$
& $\ell_{4,5}~   g^2\frac{E^2}{\Lambda^2}$
& $\ell_{8,14}~  g^4\frac{f^2_{\pi}}{\Lambda^2}$
& $\ell_{9}~    g^2\frac{E^2}{\Lambda^2}$
& $\ell_{11,12}~g^2\frac{E^2}{\Lambda^2}$\\
& & & & & & & &  \\
\hline
& & & & & & & &  \\
 $B_1^{(1)}$
& $\ell_0~ g^3\frac{f^3_{\pi}}{\Lambda^2E}$
& $\ell_{1,13}~ e^4g\frac{f^3_{\pi}}{\Lambda^2E}$
& $\ell_2~ e^2g\frac{f_\pi E}{\Lambda^2}$
& $\ell_3~ g^3\frac{f_\pi E}{\Lambda^2}$
& $\ell_{4,5}~  g^3\frac{f_{\pi}E}{\Lambda^2}$
& $\ell_{8,14}~ g^3\frac{f_{\pi}E}{\Lambda^2}$
& $\ell_{9}~    g^3\frac{f_{\pi}E}{\Lambda^2}$
& $\ell_{11,12}~g^3\frac{f_{\pi}E}{\Lambda^2}$\\
& & & & & & & &  \\
\hline\hline
\end{tabular}
\end{center}
\end{table}

\end{document}
\end